%% file: museprist.tex
\documentclass[useAMS,usenatbib]{mn2e} 

\usepackage{times,graphicx,amssymb}
\usepackage{hyperref}

\newcommand{\ion}[2]{{#1~\small#2}}

\newcommand{\HI}{$\rm H\,{\sevensize I}$}

\newcommand{\lya}{Ly$\alpha$}

\newcommand{\sbcgs}{$\rm erg~s^{-1}~cm^{-2}~\AA^{-1}~arcsec^{-2}$}
\newcommand{\sbline}{$\rm erg~s^{-1}~cm^{-2}~arcsec^{-2}$}

\newcommand{\qso}{Q0956$+$122}

\title[MUSE searches for galaxies near metal-poor LLSs]{MUSE searches for galaxies near very metal-poor gas clouds at $z \sim 3$: new constraints for cold accretion models}

\author[Fumagalli et al.]{Michele Fumagalli$^{1}$\thanks{E-mail: michele.fumagalli@durham.ac.uk}, 
  Sebastiano Cantalupo$^{2}$, Avishai Dekel$^{3}$, Simon L. Morris$^{1}$,\and 
  John M. O'Meara$^{4}$, J. Xavier Prochaska$^{5,6}$, Tom Theuns$^{1}$\\
  $^{1}$Institute for Computational Cosmology and Centre for Extragalactic Astronomy, 
  Department of Physics, Durham University, South Road, Durham, DH1 3LE, UK \\
  $^{2}$Institute for Astronomy, ETH Zurich, Wolfgang-Pauli-Strasse 27, 8093 Zurich, Switzerland\\
  $^{3}$Racah Institute of Physics, The Hebrew University, Jerusalem 91904, Israel\\
  $^{4}$Department of Chemistry and Physics, Saint Michael's College, One Winooski Park, 
  Colchester, VT 05439, USA \\
  $^{5}$Department of Astronomy and Astrophysics, University of California, 1156 High Street, 
  Santa Cruz, CA 95064 USA \\
  $^{6}$University of California Observatories, Lick Observatory 1156 High Street, Santa Cruz, 
  CA 95064 USA \\
}

\begin{document}


\pagerange{\pageref{firstpage}--\pageref{lastpage}} \pubyear{xxxx}

\maketitle

\label{firstpage}

\begin{abstract}
We report on the search for galaxies in the proximity of two very metal-poor 
gas clouds at $z\sim 3$ towards the quasar \qso. With a 5-hour MUSE integration 
in a $\sim 500\times 500$ kpc$^2$ region centred at the quasar position, we achieve a 
$\ge 80\%$ complete spectroscopic survey of continuum-detected galaxies with $m_{R} \le 25$ mag
and Ly$\alpha$ emitters with luminosity $L_{\rm Ly\alpha} \ge 3\times 10^{41} ~\rm erg~s^{-1}$. 
We do not identify galaxies at the redshift of a $z\sim 3.2$ Lyman limit system (LLS) with 
$\log Z/Z_\odot = -3.35 \pm 0.05$, placing this gas cloud in the intergalactic medium or
circumgalactic medium of a galaxy below our sensitivity limits. 
Conversely, we detect five Ly$\alpha$ emitters 
at the redshift of a pristine $z\sim 3.1$ LLS with $\log Z/Z_\odot \le -3.8$,
while $\sim 0.4$ sources were expected given the $z\sim3$ \lya\ luminosity function.
Both this high detection rate and the 
fact that at least three emitters appear aligned in projection with the LLS
suggest that this pristine cloud is tracing a gas filament that is feeding one or 
multiple galaxies. Our observations uncover two different environments for metal-poor LLSs, 
implying a complex link between these absorbers and galaxy halos, which ongoing MUSE surveys will soon 
explore in detail. Moreover, in agreement with recent MUSE observations, we detected a
$\sim 90~\rm kpc$ \lya\ nebula at the quasar redshift and three \lya\ emitters reminiscent of 
a ``dark galaxy'' population.
\end{abstract}


\section{Introduction}

The assembly and growth of galaxies throughout cosmic times requires the continuous 
accretion of substantial amounts of fresh fuel. Observations of molecular or atomic gas 
in the interstellar medium of both high-redshift systems and nearby galaxies reveal 
that, without accretion rates that are at least commensurable with the observed star 
formation rates (SFRs), galaxies would exhaust their gas reservoirs on timescales of 
$\sim 1-2~\rm Gyr$, much shorter than the Hubble time \citep[e.g.,][]{san08,gen10}. 
From this consideration, it follows that the accretion of gas from the halo (i.e. the 
circumgalactic medium or CGM) and, ultimately, from the baryon reservoir present in the 
intergalactic medium (IGM), needs to be ubiquitous at all redshifts.

In support of this argument, modern cosmological simulations predict 
accretions rates of $\gtrsim 10-20~\rm M_\odot~yr^{-1}$ at $z\sim 2-3$ 
galaxies \citep[e.g.,][]{dek09,ker09,fau11}.
However, despite a general consensus that gas accretion is a dominant process for 
galaxy evolution, direct observational evidence of cold gas infalling onto galaxies 
is scarce at $z\sim 1$ \citep{rub12,mar12} and even more tenuous at higher redshifts 
\citep[e.g.,][]{cri13,bou13,mar15}. 
While in apparent contradiction with expectations of ubiquitous inflows, the lack of 
direct detections is often justified in the context of the so-called 
``cold stream'' or ``cold flow'' paradigm if a significant fraction of the gas 
is accreted along narrow and dense filaments of cold gas \citep[e.g.,][]{bir03,ker05,dek06}. 
Indeed, if the infalling gas covers only a small fraction of the solid angle as seen 
from a galaxy \citep[e.g.,][]{goe12}, then it is natural that only a
limited number of sightlines will intersect these infalling cold streams,  
showing redshifted absorption lines. Moreover, for typical 
infall velocities of $\sim 100-200~\rm km~s^{-1}$, the signature of inflows is often 
masked by interstellar absorption at the systemic redshifts \citep{rub12}.
Further, when selecting absorbers via metal lines,
  some ambiguity remains in separating recycled gas falling back onto galaxies
  from material that is being accreted from the IGM for the first time. 

Due to these intrinsic limitations, observers have to resort to other, more
indirect, signatures of the presence of cold gas accretion. For instance, 
simulations predict that accretion in the form of cold flows is a dominant contributor 
to the cross section of optically-thick gas ($N_{\rm HI} \ge 10^{17.2}~\rm cm^{-2}$) that gives 
rise to Lyman limit systems (LLSs) near galaxies \citep[e.g.,][]{fauker11,fum11sim,van12}.
While cold streams occupy only a small fraction of the solid angle seen from 
a galaxy, the probability to intersect these filaments in the transverse direction
with background sources is higher. 
Indeed, simulations predict covering factors $f_{\rm c}$ for optically-thick 
gas within the virial radius in the range of $f_{\rm c} \sim 0.1-0.4$, although 
with large variations between different models \citep{fauker11,fum11sim,shen13,fum14cf,fau14}.
Thus, in principle, statistical comparisons between the observed properties of LLSs
near galaxies \citep[e.g.,][]{rud12,pro13} and the predictions of numerical simulations 
offer interesting constraints for the cold accretion paradigm 
\citep[see also][]{leh13,fum14cf,coo15,fum16lls}. 

On an object by object basis, however, the lack of direct kinematic signatures of infall
requires that multiple diagnostics are combined to establish whether the gas observed in absorption 
is potentially being accreted onto galaxies observed in emission at close projected 
separations. A few examples that rely on low metallicity \citep[e.g.,][]{rib11,cri13,leh13}, 
rotational signatures \citep[e.g.,][]{bou13,mar15}, or filamentary morphology \citep[e.g.,][]{can12} 
can be found in the literature. Following this approach, we present in this paper a dedicated 
search of galaxies around two very metal-poor LLSs with $Z \lesssim 5\times 10^{-4} Z_\odot$. 
Their extremely-low metallicity is at odds with what expected for gas that has been enriched by outflows.
Thus, even without direct kinematic measurements, metal-poor LLSs that reside near galaxies
are among the most compelling examples of nearly chemically-pristine gas infalling for the first time inside halos. 
It is however worth noting that very metal poor LLSs represent only a small fraction of the parent
   population, with $\lesssim 20\%$ of the LLSs having $Z \lesssim 10^{-3} Z_\odot$ between $z\sim 2.5$ and $3.5$
  \citep{fum16lls,leh16}.

Our observations target the field of the quasar  \qso\ (a.k.a. SDSSJ095852.19$+$120245.0), which 
lies at a redshift $z_{\rm qso}=3.3088\pm0.0003$ \citep{hew10} and hosts two strong absorption line systems 
along its line of sight \citep{fum11sci}: a pristine gas 
cloud at $z_{\rm lls,1}=3.096221 \pm 0.000009$ 
with \ion{H}{I} column density\footnote{Throughout this work, \HI\ 
column densities are expressed in units of cm$^{-2}$.}
$\log N_{\rm HI} = 17.18 \pm 0.04$ and without discernible metals to a limit of $Z<10^{-3.8}~\rm Z_\odot$; 
and a second LLS at $z_{\rm lls,2} = 3.223194 \pm 0.000002$, with column density $\log N_{\rm HI} = 17.36 \pm 0.05$
and metallicity $\log Z/Z_\odot = -3.35 \pm 0.05$ \citep{leh16}.
A summary of the physical properties measured for these two systems or inferred via photoionization
modelling by \citet{fum11sci} and \citet{leh16} is presented in Table \ref{tab:llsprop}.

\input{lls_table.tex}

Details of the imaging and spectroscopic observations are presented in Section \ref{sec:obs},
followed by the analysis of continuum-detected sources and \lya\ emitters in Section 
\ref{sec:continuum} and Section \ref{sec:lines}. 
In Section \ref{sec:qso}, we report on the discovery of an extended nebula at the 
quasar redshift, with discussion and conclusions in Section \ref{sec:end}.
Given the technical nature of Sections \ref{sec:obs}-\ref{sec:lines}, readers who are
primarily interested in the final results may prefer to continue reading from Section \ref{sec:qso}. 
Throughout this work, we assume solar abundances from \citet{asp09} with
$Z_\odot = 0.0134$, and we use the ``Planck 2013'' cosmology \citep{pla14} for which 
the Hubble constant is $H_0 = (67.8 \pm 0.8)~\rm km~s^{-1}~Mpc^{-1}$ and the
matter density parameter is $\Omega_{\rm m} = 0.308 \pm 0.010$. 
Magnitudes are expressed in the AB system. 

\section{Observations and Data Processing}\label{sec:obs}

\subsection{Imaging observations}

Imaging observations of the \qso\ field were obtained using LRIS \citep{oke95} at Keck, 
as part of the NOAO programmes 2013A-0078 and 2013B-0102 (PI Fumagalli). 
Thanks to the dual-arm design of LRIS, 
we acquired with the D460 dichroic a 5400 s exposure using the $u'$ filter on the blue side camera, 
together with a 1440~s exposure using the $V$ filter, and two 1080 s exposures in the $R$ 
and $I$ filters on the red side camera. 
Observations were conducted on UT 9 March 2013 when the target was transiting 
at an airmass of $\sim 1.1$ under variable conditions. 

Imaging data have been processed and calibrated following 
the procedures described in \citet{fum14dla}, including basic calibrations 
(bias and flat-fielding corrections), and photometric and astrometric calibrations.  
Due to poor seeing and patchy clouds, the resulting image quality is modest, with a
point source full-width at half-maximum (FWHM) of $1.5''$ in $u'$, $1.2''$ in 
$V$, $1.3''$ in $R$, and  $1.3''$ in $I$. 
However, we recalibrate these images using the high-quality
Sloan Digital Sky Survey (SDSS) photometry \citep{ala15} for 
bright stars in the field, achieving errors on the flux calibration within $\sim 0.06-0.05$ mag 
for all filters. This is confirmed by the excellent agreement between $R$ band magnitudes measured in LRIS images 
and images reconstructed from the Multi-Unit Spectroscopic Explorer
(MUSE) data collected for most part in photometric conditions (see below).

For each filter, we measure the noise properties and we extract source catalogues 
as described in \citet{fum14dla}. For an aperture of $2''$ in diameter, we find a 
$2\sigma$ magnitude detection limit of $27.7$ mag in $u'$, $26.9$ mag in $V$, $26.6$ mag in 
$R$, and $26.3$ mag in $I$. 

\begin{figure*}
\centering
\includegraphics[scale=1.]{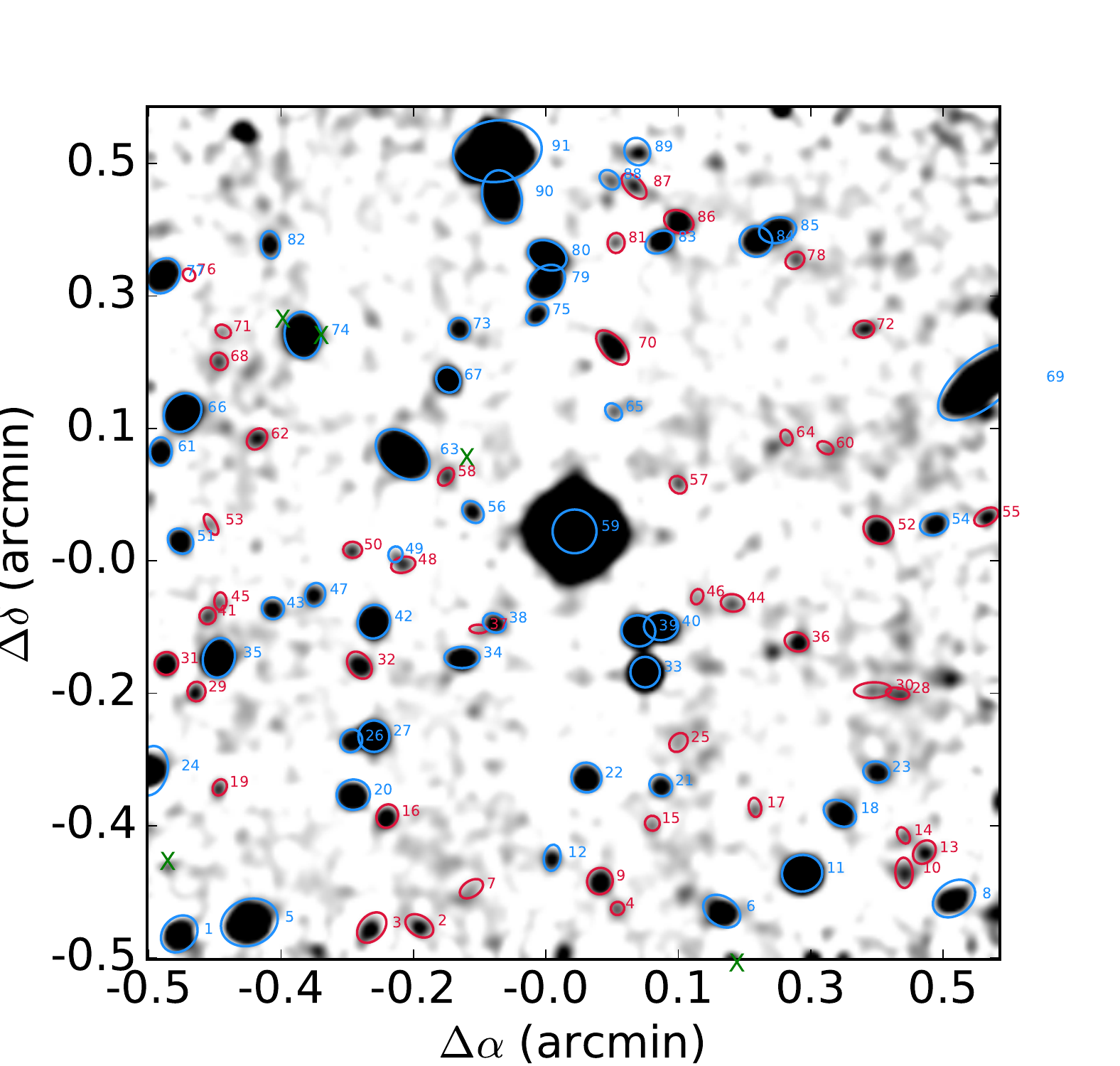}
\caption{Image of the square region ($\sim 1.1'$ on a side) centred at the position of \qso, from a deep white-light image
reconstructed by collapsing the MUSE datacube along the wavelength axis. The position of each continuum-detected source 
is marked with an aperture and the source ID. In blue, we identify sources for which a spectroscopic redshift has been 
measured, while in red we mark sources without spectroscopic redshifts. Green crosses mark \lya\ emitters
(see also Figure \ref{fig:spdis}).
Some sources are visible at the edge of the field, but they are not included in our analysis 
due to partial coverage or lower data quality in the periphery of the field of view.}\label{fovfilters}
\end{figure*}

\subsection{Spectroscopic observations}

High-resolution spectroscopy of the \qso\ quasar is available from the Keck HIRES archive,
and the analysis of this spectrum has been presented in \citet{fum11sci} and \citet{leh16}.

In addition to high-resolution spectroscopy for the quasar, we have acquired integral field unit 
(IFU) spectroscopy within a square region of $\sim 1.1'$ on a side 
centred at the quasar position. 
Observations have been conducted using MUSE
\citep{bac10} at the UT4 Very Large Telescope (VLT) as part of the programme 
094.A-0280(A) in period 94 (PI Fumagalli). Data have been acquired in Wide Field Mode
using slow guiding and the extended wavelength mode to ensure 
non-zero throughput down to 4650~\AA. A total of 4.1h on source was acquired in sets of 
1500s exposures. A dither pattern with $\sim 5''$ steps and $90~\rm deg$ rotation was adopted 
to map a $68'' \times 68''$ region around \qso\ with the $\sim 60''\times 60''$ MUSE field of view. 
With this pattern, we achieve maximum depth within the inner $40'' \times 40''$,
while in the outer region the sensitivity decreases radially from the quasar position. 
All observations were conducted in dark time, with seeing $\lesssim 0.8''$, and at 
airmass $<1.7$ (with an average of $\sim 1.3$) in clear or photometric conditions. 

Data have been reduced using a combination of recipes from the ESO MUSE pipeline \citep[v1.2.1;][]{wei14} 
and the CubExtractor package (CubEx in short, version 1.5; Cantalupo, in preparation), 
supplemented by in-house Python codes. First, we create a master bias and a master flat with the 
MUSE pipeline, together with a wavelength solution and an illumination correction using twilight flats. Next, we apply 
these calibrations to the science exposures and the standard stars used for 
spectro-photometric flux calibrations. Data cubes for individual exposures are 
constructed using a common reference grid with 
voxels of $1.25~$\AA\ in the spectral direction and $0.2''$ in the spatial direction.
At this stage, the pipeline applies a barycentric correction. 
Subsequently, using CubeFix and CubeSharp within the CubEx package, we improve the quality of the 
flat-field correction, and perform a flux-conserving sky subtraction on individual cubes 
\citep[for details see, e.g.,][]{bor16}.
The last two steps are then repeated using the cubes from the 
previous iteration to identify and mask sources in white-light images reconstructed from the cubes, 
and thus minimising the contamination from sources when computing illumination corrections.
In the end, all individual exposures are combined in a final cube using 
mean statistics. To help with source identification, we also 
construct a second cube using median statistics that better
rejects residual cosmic rays and artefacts in individual exposures.
The wavelength calibration of MUSE data is performed in air, but we apply appropriate 
transformation to vacuum when comparing to HIRES data and measuring redshifts.

Before proceeding with our analysis, we also perform a series of tests on the quality of the 
final MUSE datacube. The final depth (root mean square) of the datacube is $5\times 10^{-19}$~\sbcgs\
as measured in a 10 \AA\ window centred at 5000 \AA\ in empty regions 
near the quasar position ($4\times 10^{-19}$~\sbcgs\ at 6000~\AA).
Next, we test for the absolute flux calibration by comparing the 
$R$-band magnitudes of sources detected in a MUSE reconstructed image against values 
from our deep Keck imaging, finding a difference of $<0.05$ mag in agreement with
calibration errors. At this stage, we also verify the final image quality of the MUSE cubes,
finding a mean FWHM of $0.67''\pm0.02''$ for point sources.
We also examine the relative flux calibration and the wavelength 
solution by comparing the MUSE and SDSS spectrum for the central quasar, 
again finding excellent agreement. 

\input{source_table.tex}

\section{Properties of continuum detected sources}\label{sec:continuum}

\subsection{Source extraction and aperture photometry}

Using the MUSE median datacube, we produce a deep detection image by collapsing the cube along the 
wavelength axis. We then construct a source catalogue from the detection image by running {\sc sextractor} \citep{ber96}
with a detection area of 8 pixels and a threshold of $1.5\sigma_{\rm rms}$, where $\sigma_{\rm rms}$ is 
the background root mean square.
Only sources within a box of $65''$ on the side ($\sim 500~\rm kpc$ at the redshift of the two LLSs) 
and centred at the quasar position are considered in the following analysis, in order to avoid spurious detections 
arising from edge effects and to prevent the inclusion of sources with partial coverage.  

Next, using the segmentation map, we extract 1D spectra for each source from both the mean and median cubes, 
together with a 2D spectrum obtained by collapsing the datacube along one spatial axis. At this stage, wavelengths 
are shifted to vacuum for consistent redshift determinations in both HIRES and MUSE data.
A list of the 91 extracted sources, detailing the positions and magnitudes of each object, is 
provided in Table \ref{tab:sources}. Sources are also identified within the MUSE field of view 
in Figure \ref{fovfilters}. 

Throughout this work, we report magnitudes computed inside the Kron radius \citep{kro80}, as derived 
from the deep detection image. The same apertures, with appropriate astrometric 
transformations, are also used to extract the photometry from the Keck images. To account for the different 
image quality between the MUSE and Keck data, we apply a wavelength dependent 
aperture correction that is computed by taking the ratio of the 
seeing measurements in the Keck broadband images and the 
MUSE datacube. Agreement is found when comparing magnitudes extracted from the LRIS 
$R$ band image and an equivalent $R$ band image reconstructed from the MUSE datacube, providing a consistency 
check on both our procedures and calibrations. We emphasise, however, that while colours computed within the
MUSE datacube are very robust, colours computed across instruments are more sensitive to residual offsets in the 
astrometric calibrations.  Errors on the LRIS magnitudes are derived from the noise image following \citet{fum14dla}, 
while errors on MUSE magnitudes are computed directly from the variance cube. These errors are used 
to compute upper limits for non detections, hereafter quoted at $2\sigma$ confidence level. 

With $R$ band magnitudes for all the detected sources, we derive a simple metric
of completeness by observing that the galaxy counts reach a peak around $m_{R} \sim 26$ mag, and steeply
fall off at fainter magnitudes.

\input{emitters_table.tex}

\begin{figure*}
\centering
\includegraphics[scale=0.8]{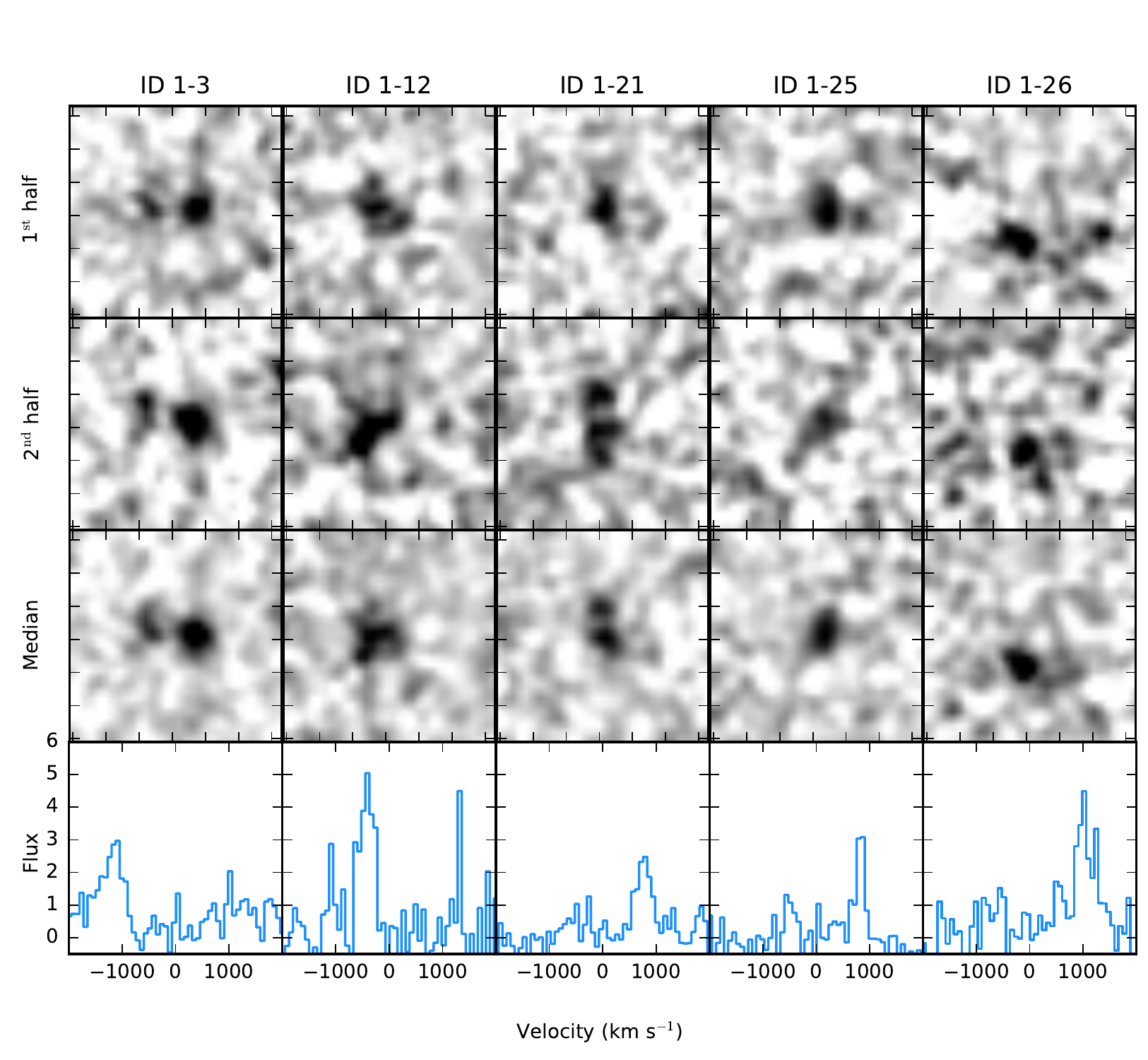}
\caption{Gallery of the five \lya\ emitters identified in proximity of the $z\sim 3.1$ LLS.
Each emitter is shown in a column, where from top to bottom we show optimally-extracted images 
obtained with CubEx \citep[Cantalupo, in preparation; see also][]{bor16}
from two independent cubes each containing half of all the exposures and from the median cube. The bottom row 
shows the 1D spectra, with velocities relative to the LLS redshift and fluxes in units of 
$10^{-20}~\rm erg~s^{-1}~cm^{-2}~\AA^{-1}$.}\label{fig:lls1}
\end{figure*}

\subsection{Spectroscopic redshifts}

To measure the redshift of the continuum detected sources, we first visually inspect each spectrum in 1D and 2D to search 
for emission and absorption lines. Weak features are confirmed by inspecting spectra from the mean and median cubes.
When multiple absorption and emission lines are identified, we determine redshifts 
from Gaussian fits to the emission lines, with typical errors of $\delta z \sim 0.0002-0.0003$. 
For sources without emission lines, instead, we compare spectra to templates from SDSS/DR5 \citep{ade07} or 
to Lyman Break Galaxy (LBG) templates from \citet{bie13} using a custom-made graphical user interface.  
For the few cases in which a single emission or absorption line 
is detected and no other strong unambiguous features are visible, we assign a tentative redshift as described in 
Table \ref{tab:sources}. Following these steps, we measure redshifts 
for $48/91$ sources ($38/91$ with robust redshifts), achieving a nearly complete spectroscopic redshift survey for sources with 
$m_{R} \le 25$ mag (with $24/28$ sources having robust redshifts). A summary of redshift measurements is 
provided in Table \ref{tab:sources}, which reveals no association with the two LLSs. 

In passing, we note the presence of a group of at least four sources at $z\sim 0.85$ (ID 22, 27, 42, 56) which is within 
$\sim 300~\rm km~s^{-1}$ from a \ion{Mg}{II} absorption system detected at $z = 0.85010 \pm 0.00001$.
Among these sources, ID 56 lies at a projected separation of $\sim 8''$ or $\sim 60~\rm kpc$ 
from the absorption line system, at nearly coincident redshift.  

\subsection{Colour selection of $z\sim 3.0-3.3$ galaxies}

For galaxies with $m_{R} > 25$ mag for which our spectroscopic redshift survey becomes severely incomplete, 
we investigate with colour information whether the MUSE field of view contains galaxies with 
spectral energy distributions (SEDs) consistent with sources at $z \sim3.0-3.3$. 
To this end, we generate observed SEDs for all the sources with continuum detection in the MUSE datacube
using six photometric data points, which are extracted from the Keck $u'$-band image 
and from five medium-band images from the MUSE datacube with a 500~\AA\ wide top-hat filter centred
at 4900~\AA, 5400~\AA, 5900~\AA, 6400~\AA, and 6900~\AA.
At this stage, we also correct the observed magnitudes for the wavelength-dependent Galactic extinction, 
following \citet{sch11}. For the broadband photometry reported in Table \ref{tab:sources}, 
these corrections are $A_{u'}=0.15$ mag and $A_{R}=0.07$ mag.

To search for sources with colours consistent with $z \sim3.0-3.3$ galaxies, we use the {\sc eazy} code 
\citep{bra08}. 
Using sources with a spectroscopic redshift as a test bed (at redshifts $z<3$), 
we find that our photometry does not span a sufficient wavelength 
range to remove the well-known degeneracy between low and high redshift sources. Moreover,
and perhaps most importantly, the $u'$-band image was acquired in modest weather conditions
thus limiting the constraining power of the (weak) upper limits from the LRIS $u'$-band image. 
This fact is indeed reflected by the broad redshift range 
spanned by the likelihood functions. For this reason, when integrating 
the redshift probability distribution function derived assuming a flat prior on the magnitude in the 
redshift interval $z=[3.0,3.3]$, we do not identify any galaxy with a significant probability excess 
($\gtrsim 50\%$) in the interval of interest. Only three sources (ID 52, 53, and 86) exhibit a modest excess
of $\sim 25-30\%$, but the broad shape of the posterior redshift distribution function prevents us from
conclusively identifying any of them as the host of one of the two LLSs.

\begin{figure}
\centering
\includegraphics[scale=0.45]{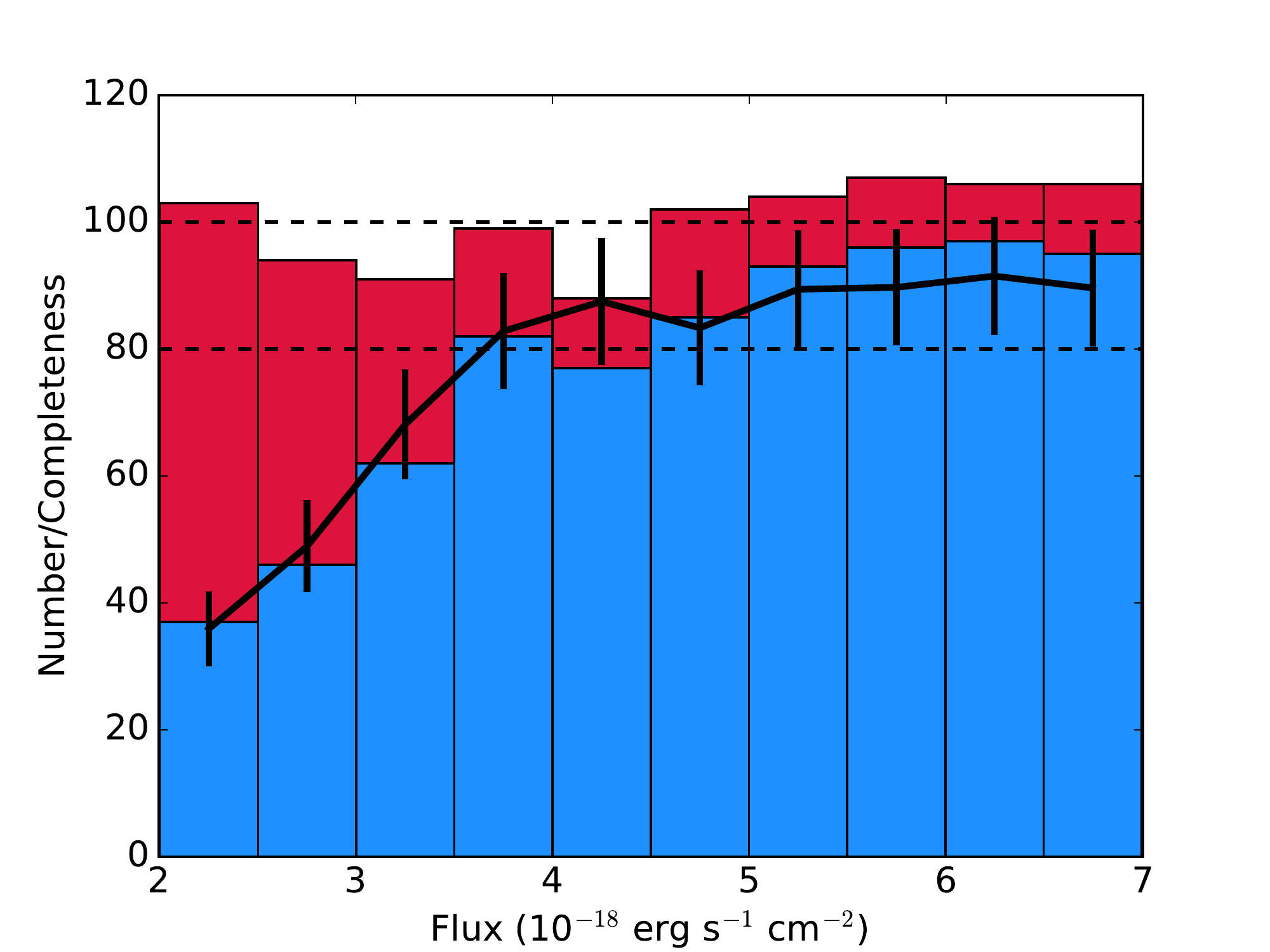}
\caption{Results of a completeness test for the recovery of \lya\ emitters between $z\sim 3.1$ and $3.2$. 
In red, we show the input mock flux distribution, and in blue we show the number of sources identified by CubEx. 
The black solid line shows the recovered fraction, including counting errors. 
Our search for \lya\ emitters is $\ge 80\%$ complete for fluxes 
$\ge 3.5\times 10^{-18}~\rm erg~s^{-1}~cm^{-2}$.}\label{fig:comp}
\end{figure}

\begin{figure*}
\centering
\includegraphics[scale=0.8]{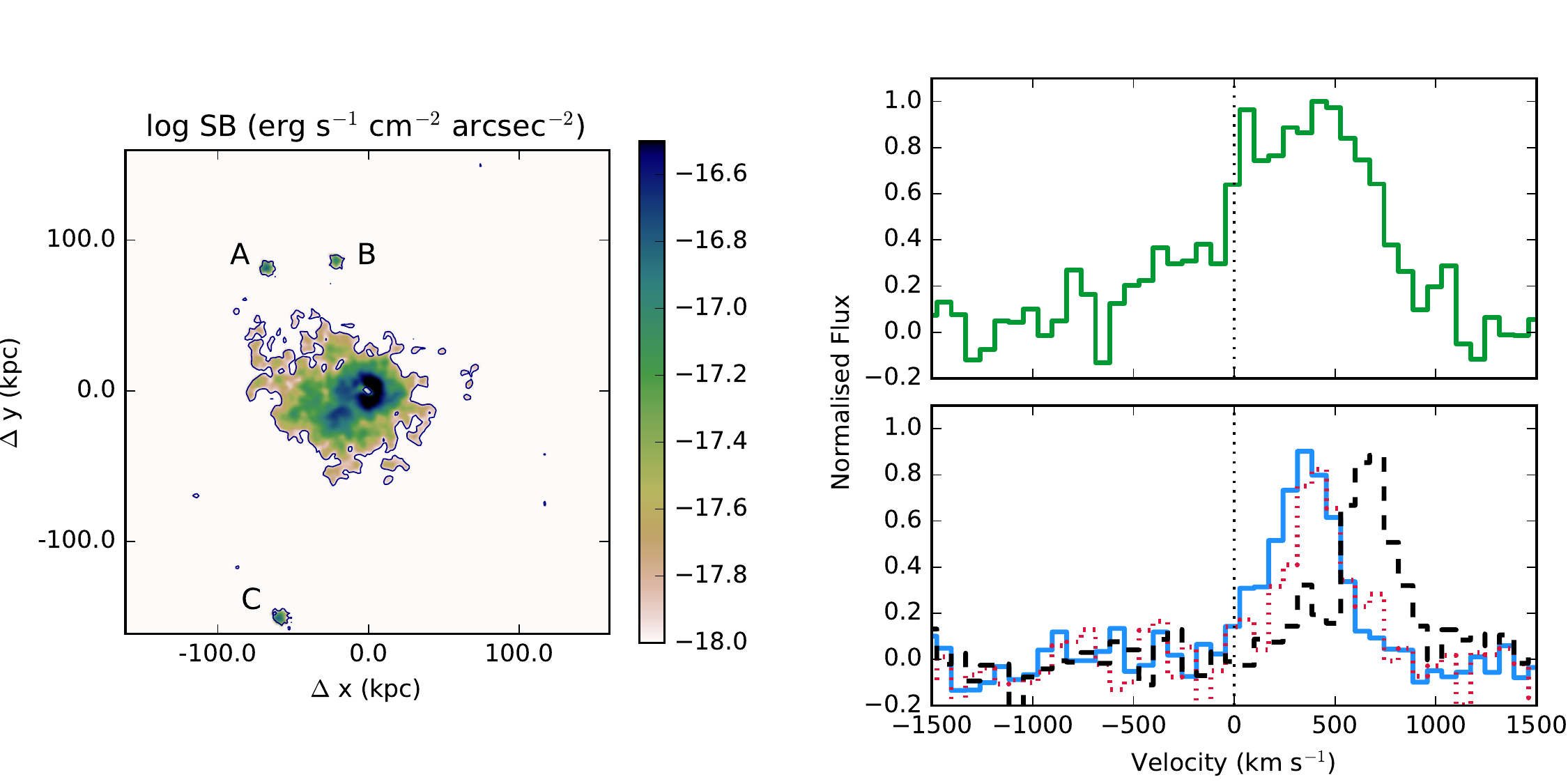}
\caption{({\it Left}) Ly$\alpha$ surface brightness map of an extended nebula and of three Ly$\alpha$ emitters discovered 
at the quasar redshift. The image shows the surface brightness derived from optimally-extracted line maps. 
The blue contour shows the $10^{-18}$ \sbline\ level. 
({\it Right}) Spectra integrated over the segmentation map for the 
extended nebula (top) and the three emitters (bottom; solid blue line for source A, red dotted line for source B, 
and black dashed line for source C). The velocity scale is centred at the quasar redshift derived from 
rest-frame UV lines (vertical dotted line).}\label{fig:nebula}
\end{figure*}

\section{Properties of line emitters}\label{sec:lines}

We focus next on the detection and characterisation of line emitters which, due to their faint
continuum level, may be undetected in the deep white-light image. 
Throughout this section, we make use of the CubEx capability (Cantalupo, in preparation) 
of identifying sources in three dimensions
by searching for groups of contiguous voxels in the MUSE datacube above a desired signal-to-noise ratio ($S/N$) 
threshold both in the spatial direction and along the wavelength axis. 

We search for line emitters in proximity of the two intervening LLSs within a window of about $\pm 1000~\rm km~s^{-1}$, 
running CubEx on the mean cube\footnote{We defer the search of emitters across the entire cube to future work 
within our ongoing Large Programme (197.A-0384).}, 
after having subtracted the quasar point spread function (PSF) and the continuum
emission of each source as described in \citet{bor16}. 
To ensure the highest completeness to faint fluxes, we extract 
candidate sources with a minimum volume of 50 voxels at $S/N \ge 3$. At this step, the cube is convolved in the spatial 
direction with a boxcar filter of 2 pixels and the variance cube is rescaled by a factor of $\sim 1.8$
(weakly dependent on wavelength) to match the value measured within the datacube. 

After having identified the sources, we visually inspect optimally-extracted line flux images \citep[for details, see][]{bor16}
both from the mean and median cube. To avoid the inclusion of spurious sources (e.g., cosmic rays residuals or image defects
arising from the edges of the frame), we also construct line flux images from two independent cubes containing only one of the two 
exposures collected in each observing block. 
Following visual inspection, a source is included in the final catalogue if it is detected both in the mean and median 
cubes, as well as in the two independent cubes constructed from half of all the exposures available. 
While this choice affects the recovery fraction at the faint end of the flux distribution, 
it ensures that only reliable sources are included in the final analysis. 

Finally, we extract 1D spectra and measure redshifts by fitting Gaussian functions to the emission lines. 
Sources are classified as Ly$\alpha$ emitters if the emission line is not resolved, which would be indicative 
of [OII] emission, and no positive identification of other emission lines (H$\alpha$, [OIII], H$\beta$) is made
in the spectrum. 

In a window of about $\pm 1000~\rm km~s^{-1}$ centred at redshift of the $z\sim 3.1$ LLS, we identify 
six line emitters, five of which are classified as Ly$\alpha$ sources. The remaining source is an [OII] emission line 
coinciding with the continuum-detected galaxy ID 18.
Conversely, at the redshift of $z\sim 3.2$ LLS, we detect only one source with more than 
50 voxels above $S/N=3$, which is a strong, spatially resolved, 
[OII] emission line associated with the galaxy ID 63 at $z \sim 0.3752$.
No other sources are identified as Ly$\alpha$ emitters within this region of the cube.

For each \lya\ emitter detected near the $z\sim 3.1$ LLS, we also measure the distance from 
the quasar centroid where \HI\ absorption is detected, and we integrate 
the line flux within the three-dimensional segmentation maps provided by CubEx. While this choice maximises the $S/N$ 
of this measurement, fluxes should be regarded as formal lower limits, although we note modest differences 
when integrating the flux in a cubic region that encompasses the full extent of the detected emission line
(see also below). A summary of the properties of the five Ly$\alpha$ emitters identified at the redshift of the $z\sim 3.1$ LLS 
is provided in Table \ref{tab:emitters} and Figure \ref{fig:lls1}.

Finally, we assess the completeness of our search for \lya\ emitters by means of mock tests. Specifically, we
populate the mean datacube (preserving the original noise) 
with line emitters with fluxes in the range $(2-7)\times 10^{-18}~\rm erg~s^{-1}~cm^{-2}$
and size defined by a two-dimensional Gaussian with $0.7''$ FWHM in the spatial direction, and a Gaussian of
2.5~\AA\ FWHM in the spectral direction. 
During this test, we inject 500 sources at the redshift of each LLS, 
which represents a compromise between reducing counting errors and avoiding blending of sources.
We then process these mock cubes following the same analysis adopted for the real data, finding similar results
for each LLS. The fraction of recovered sources combined for both LLSs is shown in Figure \ref{fig:comp},
which indicates that our search is $\ge 80\%$ complete for line fluxes $\ge 3.5\times 10^{-18}~\rm erg~s^{-1}~cm^{-2}$.
As visible in the figure, occasional blending slightly reduces the completeness also at the bright end.
At this stage, we also test the quality of the recovered line fluxes, finding that the discrepancy 
between the input and recovered fluxes is distributed as a Gaussian, the centre of which is consistent with 
zero to within the 1$\sigma$ flux errors. 

\section{An extended nebula associated with the quasar}\label{sec:qso}

Upon inspection of the datacube at wavelengths corresponding to Ly$\alpha$ at the redshift of the quasar, 
we identify extended diffuse emission as well as the presence of three compact emitters. 
After subtracting the quasar PSF and all 
continuum detected sources, we obtain a segmentation map for the diffuse emission by identifying with 
CubEx pixels with $S/N \ge 2$ inside a region of minimum area of 2000 voxels, chosen to be large enough to 
filter compact sources. 
To characterise the three emitters, we further run CubEx to search for sources composed of at least 60 voxels with $S/N\ge 5$.
Figure \ref{fig:nebula} shows the surface brightness map and 1D spectra of both the nebula and these Ly$\alpha$ emitters. 

As shown in the figure, this nebula is highly asymmetric and extends in the east direction for up to $\sim 12''$,
or $\sim 90~\rm kpc$ at the redshift of the quasar, with a surface brightness of $\ge 10^{-18}~$\sbline. 
While not as dramatic as some of the giant nebulae uncovered around $z\sim 2$ quasars \citep{can14,hen15}, 
this nebula is brighter than the typical nebulae seen around $z\sim 2$ quasars \citep{arr16}, 
and it is comparable to the nebulae seen by MUSE at $z\sim 3$ around all bright radio-quiet quasars observed 
so far \citep[e.g.,][]{bor16}. Kinematically, the bulk of the Ly$\alpha$ emission is redshifted by $\sim 500~\rm km~s^{-1}$ when considering the quasar redshift measured from rest-frame UV lines (Figure \ref{fig:nebula}, top-right panel) 
and spans $\gtrsim 600~\rm km~s^{-1}$ in velocity, with a second component 
extending to negative velocity.

The spectra of the three compact sources show prominent Ly$\alpha$ emission aligned in velocity space with the nebula
(Figure \ref{fig:nebula}, bottom-right panel). 
No obvious signatures of a strong AGN component are evident from the spectra of these 
compact sources. The emitters have a line flux integrated over the 
segmentation map of $(8.4\pm 0.4) \times 10^{-18}~\rm erg~s^{-1}~cm^{-2}$ (source A), 
$(5.1\pm 0.4) \times 10^{-18}~\rm erg~s^{-1}~cm^{-2}$ (source B), and 
$(8.9\pm 0.5) \times 10^{-18}~\rm erg~s^{-1}~cm^{-2}$ (source C).
As previously discussed, these values
represent formally lower limits, as additional flux may be present outside of the 
voxels selected by the segmentation map. However, we have shown that this effect is expected to be a minor one, 
with all the flux being recovered typically within errors. 
Source A is also detected in the continuum, with an $R$ band magnitude of $m_{R} = 27.5 \pm 0.2$ mag computed inside an 
aperture matched to the size of the \lya\ emission. However, we cannot exclude contamination from the nearby source ID 67 in 
close proximity to emitter A, and we consider this measurement an upper limit to the actual continuum. 
Conversely, source B and C are not detected in the continuum within the 
$R$ band image, to a limiting magnitude of $28.6$ mag and $28.9$ mag ($2\sigma$) within an aperture matched to the 
\lya\ emission.

With a luminosity of $L_{\rm Ly\alpha} \ge 5 \times 10^{41}~\rm erg~s^{-1}$ and lacking appreciable continuum emission,
these sources are characterised by equivalent widths $>100-140~$\AA\
in aperture matched to the extent of the \lya\ emission,
or $>150-280~$\AA\ when adopting apertures matched to the seeing of the continuum image. 
Thus, these sources are reminiscent of the population of bright emitters, including ``dark galaxies''
with equivalent widths in excess of 240~\AA, reported by \citet{can12} around a bright quasar that boosts 
the \lya\ emission intrinsic to the sources.
Indeed, in a volume of $\sim 150~\rm Mpc^3$ defined by the MUSE field of view and a velocity window of  
$\pm 1500~\rm km~s^{-1}$ around the quasar redshift, one would expect to see approximately two emitters
according to the luminosity function presented in \citet{can12} when including the boost due to
quasar radiation. This estimate is in line with our observations and a factor of $\sim 5$ above the expectation from the 
field luminosity functions \citep[e.g.,][]{cas11}. 

\begin{figure}
\centering
\includegraphics[scale=0.6]{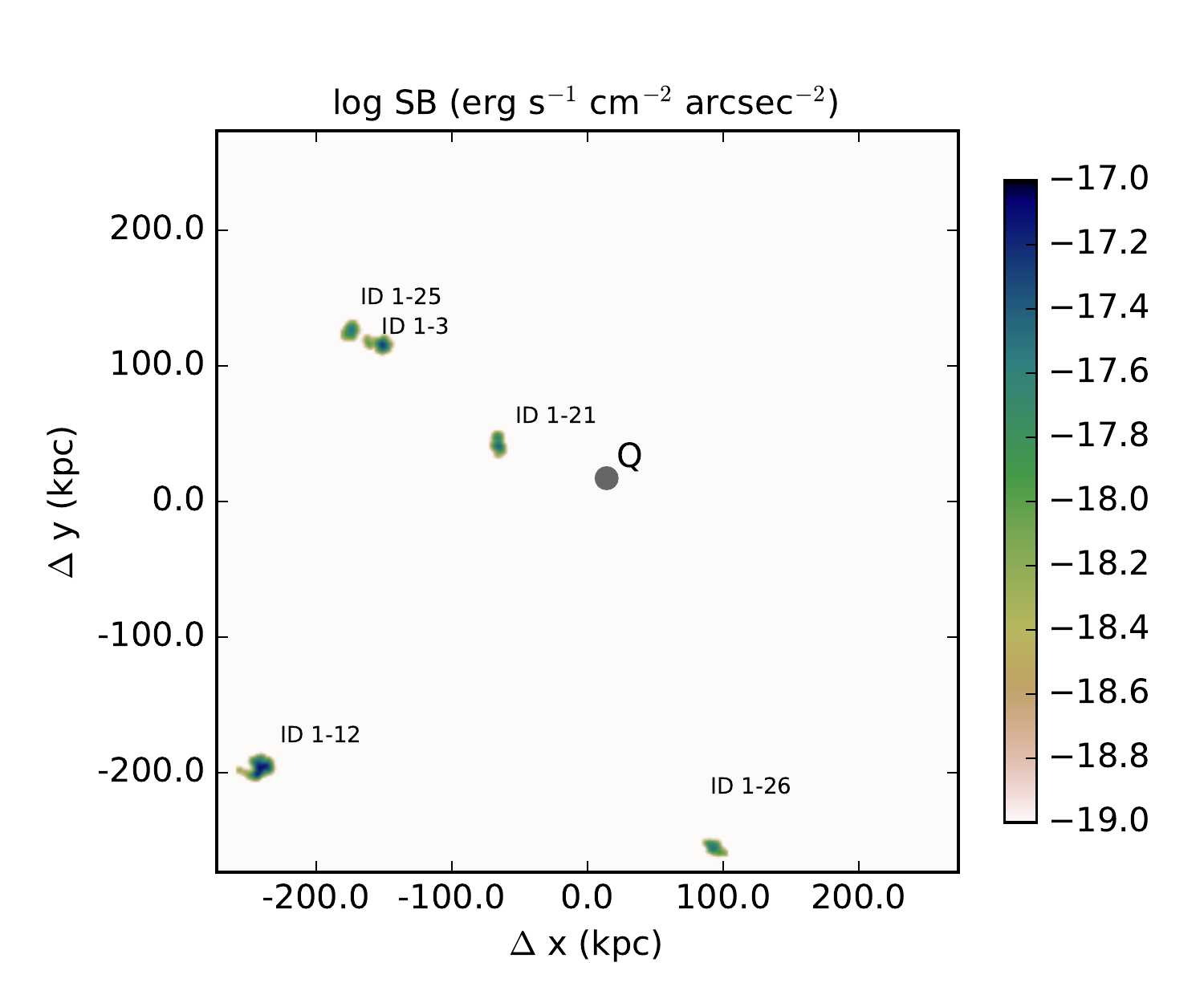}
\caption{Spatial distribution of the \lya\ emitters at the redshift of the $z\sim 3.1$ LLS.
  North is up, east to the left. The quasar position is marked by the grey circle.
  The alignment in projection of at least three 
\lya\ emitters and a pristine LLS is suggestive of a cosmic web filament crossing the field of view 
in the north-east/south-west direction.}\label{fig:spdis}
\end{figure}

\section{Discussion and Conclusions}\label{sec:end}

Our analysis of a deep MUSE exposure in the field of the quasar \qso\ that hosts two very metal-poor LLSs 
provides a proof-of-concept of the power of IFU spectroscopy at large telescopes for 
addressing open questions on the nature of LLSs and their link with the CGM of galaxies as 
predicted by simulations \citep[e.g.,][]{fauker11,fum11sim,shen13,fum14cf,fau14}.
Indeed, with a 5 hour observation, we have achieved a $\ge 80\%$ complete spectroscopic survey 
for continuum-detected galaxies with $m_{R}\le 25$ mag and for \lya\ emitters with luminosity  
$L_{\rm Lya} \ge 3 \times 10^{41}~\rm erg~s^{-1}$.
This drastically improves on what was previously possible with narrow-band or (multi-object) spectroscopic surveys 
\citep[e.g.,][]{fyn00,rau08,cri15}. 
 
In our observations, we do not identify bright continuum-detected sources at the redshift of the two LLSs,
thus excluding a rapidly star-forming galaxy (i.e. a ``classic'' LBG) 
with mean halo mass $\log M/M_\odot \sim 11.5$ \citep{bie13} and observed SFRs of 
$\gtrsim 6~\rm M_\odot~yr^{-1}$ as the host of these gas clouds. These are
the galaxies that are generally targeted in studies of galaxy-quasar pairs 
\citep[e.g.,][]{rud12,pro13}.

Furthermore, when searching for \lya\ emitters near the metal-poor ($\log Z/Z_\odot = -3.35 \pm 0.05$)
LLS at $z\sim 3.2$, we do not detect any source to a limiting luminosity of 
$L_{\rm Lya} \ge 3 \times 10^{41}~\rm erg~s^{-1}$ (uncorrected for dust or IGM absorption), 
corresponding to SFRs of $\gtrsim 0.2~\rm M_\odot~yr^{-1}$ \citep[see][]{rau08}
and for which our search is $\ge 80\%$ complete. The lack of any detection of \lya\ emitters 
within a comoving volume of $\sim 95~\rm Mpc^{-3}$ defined by the MUSE field of view and a velocity window
of $\pm 1000~\rm km~s^{-1}$ is consistent with the field luminosity function at $z\sim 3$ 
\citep{gro09,cas11} according to which $\sim 0.4$ emitters should be expected above 
our sensitivity limit\footnote{While different determinations of the 
luminosity function agree at $z\sim 3$, differences up to a factor of 2 in our estimates can arise
when choosing different parameters from the literature. However, this uncertainty does not affect our conclusions.}.
Thus, the lack of galaxies at the redshift of this metal-poor gas cloud implies that this $z\sim 3.2$ LLS arises either 
from a pocket of the IGM (possibly connecting faint galaxies) or from the CGM of a galaxy below 
  our sensitivity limit. Although it is difficult to precisely estimate the physical densities and sizes of LLSs
  \citep{fum16lls},
  we note that the IGM origin is also supported by the low density and Mpc-scale size of the
  absorbing cloud \citep[see Table \ref{tab:llsprop} and][]{leh16}.
It should be also noted that while dust is expected to play a negligible role particularly in these 
chemically pristine environments, we cannot exclude biases arising from extinction based only on our data. 

Conversely, when searching for \lya\ emitters at the redshift of the pristine ($Z<10^{-3.8}~\rm Z_\odot$)
LLS at $z\sim 3.1$, we identify five emitters (Figure \ref{fig:lls1}). 
Of those, ID 1-21 lies at a close 
impact parameter from the quasar ($D_{\rm qso} = 77.7 \pm 0.5~\rm kpc$)
at a velocity separation of  $\Delta v = 795 \pm 20 ~\rm km~s^{-1}$
with respect to the LLS redshift.
While this velocity shift is apparently high, 
redshifts of $\sim 300-400~\rm km~s^{-1}$ relative to systemic are common for \lya\ 
due to its resonant nature,
and offsets up to $\sim 800~\rm km~s^{-1}$ have been 
observed \citep{ste10,rak11}. Thus, ID 1-21 is likely to be the closest galaxy in physical 
space to the pristine LLS.
In velocity space, instead, ID 1-12 is the closest to the LLS without considering 
radiative transfer effects ($\Delta v = -375 \pm 20 ~\rm km~s^{-1}$), but at larger projected 
impact parameter ($D_{\rm qso} = 312.4 \pm 0.5~\rm kpc$).
 For comparison, most of the host galaxies of the $z\sim 1$ LLSs that have been
  reported in the literature \citep[e.g.,][]{leh13} lie
  at distances of $\lesssim 100~\rm kpc$ and velocities of $\lesssim 350~\rm km~s^{-1}$.

 Given that the physical three-dimensional distance from each source to the LLS is unknown,
 it is difficult to unambiguously identify the closest host galaxy and to directly compare to the
 results of simulations.
However, when considered altogether, the detection of five sources in a small volume centred at the 
LLS redshift is an extremely rare fluctuation compared to the field
number density of \lya\ emitters. As discussed, adopting the \citet{cas11} luminosity function,
we should expect to detect $\sim 0.4$ emitters in this volume.
Thus, the detection of five emitters corresponds to a very rare event with probability of $\sim 5\times 10^{-5}$
of occurring at random. 
For this reason, we conclude 
that the pristine LLS lies within a ``rich'' environment with a biased population of \lya\ emitters, among which 
ID 1-21 is possibly the closest association.

Moreover, as shown in Figure \ref{fig:spdis}, at least three of the detected emitters 
(four if including ID 1-26) appear to lie along a line which also intersects the location where the 
pristine LLS is detected in absorption. 
Given the lack of systematic correlation in the line-of-sight velocity of these emitters,
and given that \lya\ is a resonant transition, we cannot propose an unambiguous interpretation for this 
alignment.  However, this morphology is strongly suggestive of a filament that crosses the field in the 
north-east/south-west direction \citep[cf.][]{mol01}. This feature, in conjunction with the fact that this cloud 
has no detectable metals but lies at close separation from at least one galaxy, supports 
the idea that this LLS originates from a cold stream that connects and feeds one or multiple
\lya\ emitters with modest SFRs.
As noted above, however, the assertion that this gas is infalling is only indirectly inferred from a combination of
low metallicity and proximity to one or more galaxies, as we lack direct observational evidence that the gas
is indeed moving towards or inside a halo.
Following this argument,
even if the observed alignment of sources was coincidental and solely due to projection effects,
our data would nevertheless suggest a picture in which nearly-pristine gas is being accreted from the IGM inside
a group of emitters.

Our work adds new constraints for the scenario put forward by modern cosmological simulations,
particularly within the 
cold accretion paradigm \citep[e.g.,][]{fauker11,fum11sim,shen13,fum14cf,fau14}.
On the one hand, the association of a metal-poor LLS with multiple \lya\ emitters offers one of the most
compelling examples of chemically-pristine gas that is likely accreting onto a galaxy (or a galaxy group), 
in line with theoretical predictions. 
On the other hand, our observations uncover two different environments for these metal-poor LLSs.
This cautions against blind associations between individual very-metal poor LLSs and cold streams 
purely relying on absorption measurements.
While not ruled out, this link needs more solid footing \citep[see e.g.,][]{rib11,cri13,fum16lls}.

The fact that metal-poor LLSs should not simply be connected to accretion 
based on their metal content alone is also reinforced by considerations on the \HI\ column density.
Indeed, besides having very low metal content, both LLSs targeted by our observations also have a 
relatively low column density ($\log N_{\rm HI} = 17.2-17.4$), at the limit of the threshold that defines 
optically-thick absorbers. 
With the exception of some partial LLSs, studies of large samples of absorbers
  at $z\sim 3$ hint at a decline in the metal
distribution of optically-thick gas at low column densities and low physical densities
\citep[e.g.,][]{coo15,fum16lls,gli16,leh16}, implying
that metal poor absorption line systems around $\log N_{\rm HI} \sim 17$
may arise not only from CGM gas, but also from the IGM.
Such a contribution from both CGM and IGM in low-column density and metal-poor LLSs
would explain the different environments seen around these two LLSs, a conclusion which is
also reinforced by the different densities and sizes inferred for these two absorbers (Table \ref{tab:llsprop}).

In summary, while we cannot draw far-reaching conclusions on the nature of LLSs and their 
connection to the CGM of galaxies from a single field, our observations place these two very metal poor LLSs
in an environment that resembles the IGM or the CGM of galaxies with very modest SFRs in one case, 
and in an  environment that is consistent with a cold stream feeding one or more galaxies in the second case.
While our results should not be trivially extrapolated to 
the full LLS population because very metal poor systems represent only $\lesssim 20\%$ of the LLS population at these redshifts
  \citep{fum16lls,leh16},
our analysis provides a clear indication that 
the claimed connection between metal-poor LLSs and star-forming galaxies fed by cold streams
is plausible, but still requires empirical scrutiny across a wide range of metallicity and column density. 
The answer to this open question is likely within reach in the era of large-field IFUs at 8m telescopes, thanks to 
dedicated observations (such as our own MUSE programme 197.A-0384(A)) that will soon 
target $z\sim 3.5$ quasar fields hosting $z\sim 3$ LLSs.
Combined with new observations at lower redshift, these IFU surveys will also
  enable detailed comparisons with studies at $z\sim 1$ 
  that currently place most of the optically-thick absorbers
  within $\sim 100~\rm kpc$ of galaxies regardless to their metallicity
  \citep[e.g.,][]{leh13}.

\section*{Acknowledgements}
We thank M. Fossati for useful discussion on the analysis of MUSE data and N. Lehner and C. Howk 
for sharing the analysis of the $z\sim 3.2$ LLS prior to publication. 
We thank J. Hennawi for his contribution to the preparation of the MUSE observing proposal. 
MF acknowledges support by the Science and Technology 
Facilities Council [grant number  ST/L00075X/1]. 
This work is based on observations collected at the European Organisation for Astronomical Research 
in the Southern Hemisphere 
under ESO programme ID 094.A-0280(A). 
Some of data presented herein were obtained at the W.M. Keck Observatory, which is operated as a scientific partnership
among the California Institute of Technology, the University of California and the National Aeronautics and Space 
Administration. The Observatory was made possible by the generous financial support of the W.M. Keck Foundation.
Keck telescope time was granted by NOAO, through the Telescope 
System Instrumentation Program (TSIP), funded by NSF.
We acknowledge the very significant cultural role that the summit of Mauna Kea has always had within the 
indigenous Hawaiian community.  We are most fortunate to have the opportunity to conduct observations from 
this mountain. This research made use of Astropy, a community-developed core Python package for Astronomy \citep{astropy}.
For access to the data and codes used in this work, please contact the authors or visit 
\url{http://www.michelefumagalli.com/codes.html}. 


\label{lastpage}

\end{document}

%% file: lls_table.tex
\begin{table*}
  \caption{Physical properties of the two LLSs, which are measured in absorption or inferred via photoionization
    modelling.}\label{tab:llsprop} 
\centering
\begin{tabular}{c c c c c c c c}
\hline
\hline
ID & $z_{\rm abs}$ & $\log N_{\rm HI}$ ($\rm cm^{-2}$) & $\log Z/Z_\odot$ & $\log n_{\rm H}$ ($\rm cm^{-3}$) & $\log x_{\rm HI}$ & $\log \ell$ (pc) \\
\hline
LLS 1    & $3.096221 \pm 0.000009$    &  $17.18 \pm 0.04$  & $<-3.8$            & $<-2.0$     &$<-2.4$      & $>3.1$ \\
LLS 2    & $3.223194 \pm 0.000002$    &  $17.36 \pm 0.05$  & $-3.35 \pm 0.05$   & $-3.3$      &$-4.1$       & 6.3   \\
\hline
\hline   
\end{tabular}
\flushleft{The columns of the table are: (1) the system ID; (2) the redshift measured in absorption;
  (3) the neutral hydrogen column density measured in absorption; (4) the inferred metallicity; (5) the inferred neutral fraction;
  (6) the inferred size. Values are from \citet{fum11sci} and \citet{leh16}.}
\end{table*}

%% file: source_table.tex
\begin{table*}
\caption{Properties of the sources with continuum detection.}\label{tab:sources} 
\centering
\begin{tabular}{c c c c c c c c}
\hline
\hline
ID & R.A. (deg) & Dec. (deg) & $m_{R}$ (mag) & $m_{u'}$ (mag)& $D_{\rm qso}$ (arcsec) & $z_{\rm spec}$ &Notes\\
\hline
1&149.72597& 12.03741 &23.91$\pm$0.12 &25.02$\pm$0.19  & 42.6 &0.6502&\\		  
2&149.72082& 12.03758 &26.13$\pm$0.40 &$>$26.11        & 32.0 &-&\\			
3&149.72184& 12.03755 &25.80$\pm$0.30 &$>$26.65        & 33.6 &-&\\			
4&149.71657& 12.03795 &26.82$\pm$0.51 &$>$26.64        & 28.6 &-&\\			
5&149.72446& 12.03765 &23.02$\pm$0.09 &23.75$\pm$0.12  & 38.4 &0.9456&\\		
6&149.71433& 12.03789 &24.60$\pm$0.19 &25.26$\pm$0.25  & 30.6 &0.8436&\\		
7&149.71970& 12.03836 &27.33$\pm$0.69 &$>$26.65        & 28.0 &-&\\			
8&149.70935& 12.03815 &24.70$\pm$0.21 &24.84$\pm$0.28  & 39.7 &1.0306&\\		
9&149.71694& 12.03852 &25.53$\pm$0.26 &$>$26.65        & 26.4 &-&\\			
10&149.71042& 12.03869&26.11$\pm$0.35 &$>$26.65        & 35.7 &-&\\			
11&149.71261& 12.03869&22.63$\pm$0.07 &23.42$\pm$0.03  & 30.9 &0.8177&\\		
12&149.71797& 12.03901&26.07$\pm$0.34 &$>$26.65        & 24.6 &1.2193*& Sky residuals\\ 
13&149.70999& 12.03913&26.38$\pm$0.40 &$>$26.65        & 35.7 &-&\\			
14&149.71043& 12.03948&26.52$\pm$0.44 &$>$26.65        & 33.7 &-&\\			
15&149.71582& 12.03973&26.82$\pm$0.51 &$>$26.65        & 22.7 &-&\\			
16&149.72151& 12.03988&25.69$\pm$0.28 &27.30$\pm$0.71  & 25.7 &-&\\			
17&149.71362& 12.04007&26.29$\pm$0.38 &$>$26.65        & 24.8 &-&\\			
18&149.71180& 12.03994&24.88$\pm$0.19 &$>$26.64        & 29.1 &0.3391&\\		
19&149.72510& 12.04049&26.44$\pm$0.39 &$>$26.65        & 33.1 &-&\\			
20&149.72223& 12.04033&24.63$\pm$0.17 &25.77$\pm$0.22  & 26.0 &0.2555*& [OIII]+MgI?\\	
21&149.71564& 12.04053&25.79$\pm$0.30 &$>$26.65        & 20.1 &0.5084&\\		
22&149.71723& 12.04069&24.45$\pm$0.16 &$>$26.65        & 18.5 &0.8518&\\		
23&149.71102& 12.04081&25.78$\pm$0.31 &26.62$\pm$0.45  & 29.0 &1.0565&\\		
24&149.72658& 12.04083&24.00$\pm$0.13 &25.43$\pm$0.47  & 36.8 &1.0058*& Sky residuals\\ 
25&149.71526& 12.04143&27.06$\pm$0.59 &$>$26.65        & 17.6 &-&\\			
26&149.72228& 12.04146&25.58$\pm$0.27 &26.15$\pm$0.30  & 23.1 &0.0485&\\		
27&149.72179& 12.04156&23.60$\pm$0.11 &26.05$\pm$0.28  & 21.7 &0.8516&\\		
28&149.71056& 12.04246&26.54$\pm$0.48 &26.38$\pm$0.36  & 27.2 &-&\\			
29&149.72560& 12.04250&26.49$\pm$0.43 &$>$26.65        & 31.1 &-&\\			
30&149.71110& 12.04253&26.68$\pm$0.49 &27.30$\pm$0.71  & 25.4 &-&\\			
31&149.72624& 12.04309&25.23$\pm$0.23 &25.67$\pm$0.27  & 32.5 &-&\\			
32&149.72210& 12.04305&25.73$\pm$0.29 &26.73$\pm$0.47  & 19.2 &-&\\			
33&149.71597& 12.04290&22.46$\pm$0.06 &26.46$\pm$0.38  & 11.8 &0.0000& Star\\		
34&149.71990& 12.04321&25.09$\pm$0.22 &26.98$\pm$0.57  & 12.8 &0.0439*& H$\alpha$?\\	
35&149.72512& 12.04320&23.69$\pm$0.11 &24.14$\pm$0.05  & 28.6 &1.0847&\\		
36&149.71273& 12.04354&26.08$\pm$0.35 &$>$26.65        & 18.6 &-&\\			
37&149.71955& 12.04381&27.24$\pm$0.63 &25.91$\pm$0.25  & 10.4 &-&\\			
38&149.71921& 12.04394&25.99$\pm$0.33 &27.39$\pm$0.75  &  9.2 &1.2003&\\		
39&149.71612& 12.04377&23.46$\pm$0.10 &23.72$\pm$0.04  &  8.8 &1.0091&\\		
40&149.71563& 12.04386&23.69$\pm$0.11 &23.61$\pm$0.03  &  9.6 &1.0090&\\		
41&149.72536& 12.04408&26.54$\pm$0.44 &26.49$\pm$0.39  & 28.5 &-&\\			
42&149.72179& 12.04396&23.83$\pm$0.12 &25.38$\pm$0.16  & 16.7 &0.8520&\\		
43&149.72396& 12.04424&25.83$\pm$0.31 &$>$26.65        & 23.6 &1.1315&\\		
44&149.71410& 12.04436&26.41$\pm$0.40 &$>$26.64        & 13.0 &-&\\			
45&149.72508& 12.04439&27.15$\pm$0.62 &$>$26.65        & 27.3 &-&\\			
46&149.71486& 12.04449&26.94$\pm$0.53 &$>$26.65        & 10.4 &-&\\			
47&149.72305& 12.04453&25.85$\pm$0.31 &26.78$\pm$0.48  & 20.2 &1.0862*&Sky residuals\\  
48&149.72116& 12.04515&26.31$\pm$0.36 &$>$26.65        & 13.3 &-&\\			
49&149.72133& 12.04538&26.88$\pm$0.46 &$>$26.65        & 13.7 &1.1574&\\		
50&149.72225& 12.04547&26.32$\pm$0.36 &$>$26.66        & 16.9 &-&\\			
\hline   
\end{tabular}
\flushleft{The columns of the table are: (1) the source ID; (2,3) the source position (J2000); 
(4) the $R$ band magnitude from MUSE data; (5) the $u'$ band magnitude from LRIS data; 
(6) the distance from the quasar;  (7) the spectroscopic redshift. For non detections, we quote $2\sigma$ limits.
Here we report observed magnitudes not corrected for galactic extinction.
Asterisks mark uncertain redshift determinations due to single emission/absorption lines, 
or identification of absorption lines in low $S/N$ spectra or next to sky lines.}
\end{table*}

\begin{table*}
\contcaption{Properties of the sources with continuum detection.}
\centering
\begin{tabular}{c c c c c c c c}
\hline
\hline
ID & R.A. (deg) & Dec. (deg) & $m_{R}$ (mag) & $m_{u'}$ (mag)& $D_{\rm qso}$ (arcsec) & $z_{\rm spec}$ & Notes\\
\hline
51&149.72594& 12.04565& 25.14$\pm$0.22 &25.82$\pm$0.23  & 29.9&0.8051&\\
52&149.71097& 12.04589& 25.39$\pm$0.25 &$>$26.65        & 22.8&-&\\
53&149.72528& 12.04600& 26.74$\pm$0.48 &$>$26.65        & 27.6&-&\\
54&149.70978& 12.04600& 25.34$\pm$0.24 &$>$26.65        & 27.1&4.0520&\\
55&149.70866& 12.04616& 26.17$\pm$0.36 &26.43$\pm$0.37  & 31.0&-&\\
56&149.71967& 12.04626& 26.18$\pm$0.35 &$>$26.65        &  7.9&0.8506&\\
57&149.71527& 12.04683& 26.89$\pm$0.51 &$>$26.65        &  8.5&-&\\
58&149.72025& 12.04700& 26.21$\pm$0.36 &$>$26.65        & 10.7&-&\\
59&149.71749& 12.04586& 17.16$\pm$0.01 &20.74$\pm$0.00  &  0.0&3.3088& Quasar\\
60&149.71211& 12.04761& 26.90$\pm$0.59 &$>$26.43        & 19.9&-&\\
61&149.72636& 12.04753& 25.08$\pm$0.20 &26.90$\pm$0.55  & 31.9&0.2668&\\
62&149.72430& 12.04779& 26.40$\pm$0.43 &$>$26.65        & 25.1&-&\\
63&149.72117& 12.04747& 23.10$\pm$0.09 &24.43$\pm$0.07  & 14.3&0.3752&\\
64&149.71294& 12.04782& 26.87$\pm$0.53 &$>$26.65        & 17.4&-&\\
65&149.71665& 12.04836& 27.58$\pm$0.65 &$>$26.65        &  9.5&1.5572&\\
66&149.72590& 12.04835& 22.92$\pm$0.08 &24.15$\pm$0.07  & 31.1&0.5491&\\
67&149.72020& 12.04903& 24.23$\pm$0.14 &$>$26.65        & 15.0&0.3554&\\
68&149.72511& 12.04942& 26.92$\pm$0.56 &$>$26.65        & 29.9&-&\\
69&149.70876& 12.04899& 21.70$\pm$0.05 &24.16$\pm$0.18  & 32.7&0.2710&\\
70&149.71668& 12.04971& 25.12$\pm$0.21 &27.19$\pm$0.65  & 14.2&-&\\
71&149.72502& 12.05005& $>$26.78       &$>$26.65        & 30.7&-&\\
72&149.71128& 12.05010& 26.29$\pm$0.38 &$>$26.65        & 26.6&-&\\
73&149.71996& 12.05011& 25.74$\pm$0.29 &$>$26.65        & 17.7&0.3336&\\
74&149.72332& 12.04997& 23.02$\pm$0.08 &24.46$\pm$0.10  & 25.5&0.5556&\\
75&149.71829& 12.05040& 25.86$\pm$0.29 &$>$26.65        & 16.7&0.0740*& H$\alpha$?\\
76&149.72576& 12.05124& $>$27.39       &$>$26.65        & 35.1&-&\\
77&149.72631& 12.05122& 23.88$\pm$0.12 &26.01$\pm$0.29  & 36.7&0.3555*& Template match\\
78&149.71276& 12.05154& 26.78$\pm$0.52 &$>$26.65        & 26.4&-&\\
79&149.71809& 12.05108& 23.88$\pm$0.12 &24.78$\pm$0.13  & 19.0&2.1311&\\
80&149.71807& 12.05165& 23.47$\pm$0.10 &24.99$\pm$0.13  & 21.0&0.5850&\\
81&149.71660& 12.05191& 26.65$\pm$0.43 &$>$26.65        & 22.1&-&\\
82&149.72401& 12.05187& 25.94$\pm$0.33 &26.65$\pm$0.44  & 31.7&1.2034&\\
83&149.71566& 12.05193& 25.22$\pm$0.23 &25.30$\pm$0.15  & 22.8&2.2912&\\
84&149.71360& 12.05194& 24.64$\pm$0.17 &25.14$\pm$0.13  & 25.8&0.5094&\\
85&149.71313& 12.05217& 24.98$\pm$0.22 &24.72$\pm$0.12  & 27.4&0.8691*& MgII?\\
86&149.71525& 12.05235& 25.36$\pm$0.25 &$>$26.64        & 24.7&-&\\
87&149.71621& 12.05310& 26.65$\pm$0.45 &$>$26.65        & 26.5&-&\\
88&149.71673& 12.05323& 27.09$\pm$0.60 &27.26$\pm$0.68  & 26.8&0.1401*& H$\alpha$?\\
89&149.71614& 12.05382& 26.00$\pm$0.33 &$>$26.65        & 29.1&4.1196*& Ly$\alpha$?\\
90&149.71905& 12.05287& 22.77$\pm$0.08 &23.34$\pm$0.08  & 26.0&0.6230&\\
91&149.71914& 12.05383& 21.10$\pm$0.04 &22.88$\pm$0.05  & 29.4&0.5562&\\
\hline    						
\end{tabular}
\end{table*}

%% file: emitters_table.tex
\begin{table*}
\caption{Properties of the Ly$\alpha$ emitters near to the pristine $z\sim 3.1$ LLS.}\label{tab:emitters} 
\centering
\begin{tabular}{c c c c c c c}
\hline
\hline
ID & R.A. (deg) & Dec. (deg) &  $\Delta v\rm (km~s^{-1})$ & D$_{\rm qso}$ (kpc,$"$) & $F_{\rm Ly\alpha}\rm (10^{-18} erg~s^{-1} cm^{-2})$ \\
\hline
1-3      &149.72298&12.04986& -1149  &189.5, 24.26   &$ 5.9\pm  0.5$ \\
1-12     &149.72627&12.03883&  -375  &312.4, 40.00   &$13.7\pm  1.1$ \\
1-21     &149.71985&12.04730& 795    & 77.7,  9.95   &$ 4.0\pm  0.4$ \\
1-25     &149.72380&12.05021& 861    &213.4, 27.32   &$ 3.4\pm  0.4$ \\
1-26     &149.71406&12.03670&  1033  &273.4, 35.01   &$ 3.6\pm  0.6$ \\
\hline
\hline   
\end{tabular}
\flushleft{The columns of the table are: (1) the source ID; (2,3) the source position; 
(4) the relative velocity along the line of sight with respect to the LLS, with typical errors of $\sim 15-20~\rm km~s^{-1}$;
(5) the projected distance relative to the quasar centroid where \HI\ absorption is measured, with typical errors 
of $\sim 0.5$ kpc or $\sim 0.06"$; (6) the line flux integrated within the CubEx segmentation map.}
\end{table*}